# Towards a Semantics of Activity Diagrams with Semantic Variation Points

Hans Grönniger[1], Dirk Reiß[2], Bernhard Rumpe[1]

[1] Software Engineering, RWTH Aachen University, Germany
[2] Institut für Wirtschaftsinformatik, Abteilung Informationsmanagement
Technische Universität Braunschweig, Braunschweig, Germany

**Abstract.** UML activity diagrams have become an established notation to model control and data flow on various levels of abstraction, ranging from fine-grained descriptions of algorithms to high-level workflow models in business applications. A formal semantics has to capture the flexibility of the interpretation of activity diagrams in real systems, which makes it inappropriate to define a fixed formal semantics. In this paper, we define a semantics with semantic variation points that allow for a customizable, application-specific interpretation of activity diagrams. We examine concrete variants of the activity diagram semantics which may also entail variants of the syntax reflecting the intended use at hand.

## 1 Introduction

Activity diagrams [1] are a widely accepted modeling language for representing control and data flow within software systems. The notation is applicable to various application domains and is useful on many levels of abstraction. To name just a few forms of use, activity diagrams can be used for low-level descriptions of algorithms similar to flow-charts [2], for modeling collaborating objects in an object-based system, or for specifying simple web application page flows [3] and high-level business application workflows [4].

The basic idea of activity diagrams is to model actions and their possible orders of execution. Besides this common denominator, interpretation of what constitutes an action and how to determine when and how an action is enabled or when it finishes execution remains specific to the application area. Methodically, the purpose of activity diagrams is also subject to project-specific interpretation: it may be loosely used for documentation purposes, or formally employed for analysis or code generation.

Formal semantics for activity diagrams helps to reduce misunderstandings between people and may enhance interoperability between tools. Because of the flexibility of the notation regarding its possible forms of use, it turned out to be inappropriate to use a single and fixed formal semantics. Instead, we define a semantics with semantic variation points which allow for a customizable, application-specific interpretation of activity diagrams. Explicit semantic variation points help people to agree on the meaning of language constructs in a certain project context. Invariant definitions constitute what we call the *inner semantics* of a notation. This separation helps to reduce the complexity of agreeing on a formal semantics. This paper concentrates on defining the inner semantics. Additionally, variants of activity diagrams, for example, to model low-level algorithms, are sketched.

The paper is structured as follows. In Section 2, we shortly describe the concrete and abstract syntax of activity diagrams. In Section 3, we define a formal inner semantics with variation points, which are interpreted in the different contexts in Section 4. In Section 5, we discuss related work and Section 6 concludes the paper.

## 2  Syntax of Activity Diagrams

Fig. 1 shows an example activity diagram. The workflow depicted therein describes an abstract view on a process for grading a thesis. It involves three roles (denoted on the left hand side): Student, Referee1 and Referee2. The workflow starts with a student who files a thesis. The action FileThesis has an output pin (Thesis t) that represents type and name of the outgoing data. The thesis is reviewed by Referee1 and Referee2 (fork to actions ReviewThesis1 *and* ReviewThesis2). Both actions have input and output pins – taking a Thesis t as input and passing on a Review r along the flow. When both actions have finished, the reviews are then evaluated by Referee1 (action Evaluate). Depending on the outcome of this action, either a certificate for the student is created (action CreateCert in case of passed) or note of the failure (action DetainFailure in case of failed) is taken. After either action, the activity is finished.

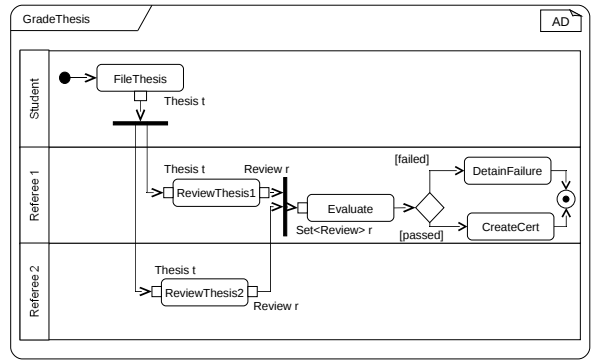

**Fig. 1.** Example activity "GradeThesis"

Please note that we currently do not consider constructs like hierarchical decomposition, interruptible activity regions or parameter sets which are present in the UML 2.2 [1] standard. Additional constructs can and will be handled in subsequent versions of the semantics. The focus of this work, however, is to show the handling of variants for the interpretation of activity diagrams.

### 2.1 Abstract Syntax

The abstract syntax of an activity diagram is given in Definition 1. An activity diagram has a name, a set of nodes and transitions. roleOf associates a role (being a name) to each node. A Node has a type, a name, a list of input and output pins, and some Effect when executed which remains unspecified for now. Transitions connect input and output pins of nodes. The names Src, Dst refer some node, the names InPin, OutPin refer to pins of the connected nodes. pinType yields the (data) type of the pin which also remains unspecified. Guards may be specified on outgoing transitions but actually belong to the source node. Therefore, we associate guards with output pins. These guards can be obtained by function guard. The exact structure of the language of guards is also not fixed in Definition 1.

**Definition 1 (Abstract Syntax of Activity Diagrams).**

$$
\begin{aligned}
&\text{AD} &&= \text{Name} \times \wp(\text{Node}) \times \wp(\text{Transition}) \times \\
& && \quad \text{roleOf} \times \text{pinType} \times \text{guard} \\
&\text{Node} &&= \text{NType} \times \text{NName} \times \text{InPin}^* \times \text{OutPin}^* \times \text{Effect} \\
&\text{roleOf} &&\in \text{Node} \to \text{Role} \\
&\text{NType} &&= \{\text{action}, \text{initial}, \text{final}, \text{forkjoin}, \text{decisionmerge}\} \\
&\text{Transition} &&= \text{Src} \times \text{OutPin} \times \text{Dst} \times \text{InPin} \\
&\text{guard} &&\in \text{PName} \to \text{Guard} \\
&\text{InPin}, \text{OutPin} &&= \text{PName} \\
&\text{pinType} &&\in \text{PName} \to \text{PType} \\
&\text{Src}, \text{Dst} &&= \text{NName} \\
&\text{Role}, \text{NName}, \text{PName} &&= \text{Name}
\end{aligned}
$$

As can be seen from Definition 1 the usually distinct node types for fork and join as well as decision and merge have been combined to more general nodes. A single fork, for example, is just a special case of a node of type forkjoin with exactly one input transition.

We introduce helper functions that operate on the abstract syntax for convenience.

- inT : AD × Node → $\wp$(Transition) yields all incoming transitions given a node.
- outT : AD × Node → $\wp$(Transition) returns all outgoing transitions of a node.
- Dot-notation is used to access parts of the abstract syntax. For instance, if $ad \in \text{AD}$, then $ad.\text{Node}$ denotes the set of nodes in $ad$.

Further, we assume that the following context conditions hold (among others). The diagram is complete in the sense that all nodes define pins when connected by a transition. Pins that are only control pins are given the (pseudo) type $\bot$. Pins with an underspecified data type are given the type $\top$ representing arbitrary values. Each transition references existing nodes and pins. In the concrete syntax (cf. Fig. 1), pins and their types may be left out but are assumed to be present in the abstract syntax.

As for the concrete syntax, we only consider a true subset of constructs compared to the UML standard. We also refrain from defining a simplified metamodel for the abstract syntax because our set-based notation is more succinct, precise, and convenient when defining the semantic mapping.

## 3 Inner Semantics of Activity Diagrams

We give a denotational semantics to activity diagrams. To do so, we precisely and explicitly define the (abstract) syntax (previous section), the semantic domain, and the semantic mapping [5].

### 3.1 System Model

The system model in the form of [6, 7] serves as our semantic domain. It characterizes object-based systems by describing their structural, behavioral, and interaction aspects. The purpose of the system model is to have a common semantic domain for all kinds of UML diagram types. As described in [8] several UML sub-languages have already been mapped to the system model. A set-valued semantic mapping for individual diagram types allows for integrating multiple semantics: the integrated semantics of a set of models denotes all systems in the system model that fulfill all properties induced by the models. Object references are available as elements of a *universe of object identifiers* UOID[3]. Similarly, a universe of class names (UCLASS), variable names (UVAR), values (UVAL), methods (UMETH), threads (UTHREAD), and program counters (UPC) is part of each system in the system model providing static information. For each object $oid$, $\text{classOf}(oid) \in \text{UCLASS}$ determines its class. All methods $m$ are defined in a class: $\text{definedIn}(m) \in \text{UCLASS}$ and there is a set of program counters for each method, i.e., $\text{pcOf}(m) \subseteq \text{UPC}$.

From a global view-point, a system of the system model is a single non-deterministic state machine. The behavior is determined by a transition function of the form

$$\Delta : \text{STATE} \to \wp(\text{STATE})$$

where STATE is the set of global states. Each state $s \in \text{STATE}$ consists of three components. The data store $\text{dsOf}(s) \in \text{UOID} \to (\text{UVAR} \to \text{UVAL})$ of a state $s$ captures attribute values of all currently existing objects. The control store $\text{csOf}(s) \in \text{UOID} \to \text{UTHREAD} \to \text{Stack}(\text{FRAME})$ saves computational states of methods in a stack of frames for each object and thread. A frame $f = (callee, mname, vars, pc, caller) \in \text{FRAME}$ stores the called object reference, the method name, current local variables, the current program counter, and the calling object. To access the program counter, we define $\pi_{pc}(f) = pc$. Finally, the event store $\text{esOf}(s)$ holds unprocessed messages. As can be seen from the transition function above, we have a closed-world assumption. Inter-object communication is hidden in the global view since messages are sent directly to the receiving event stores. Concurrent activities are possible in one state transition because a global state as a whole captures the individual state of each object. A trace $t \in \text{TRACE}$ of a system in the system model is a finite or infinite sequence of states

$$t = s_1 \cdot s_2 \cdot s_3 \cdots \text{ such that } s_{i+1} \in \Delta(s_i)$$

---

[3] All elements are defined in the context of a system $sm \in \text{SystemModel}$. We write UOID but actually refer to a specific system's set of object identifiers $\text{UOID}_{sm}$.

For details regarding the rationale behind the system model and the actual definitions please consult [6, 7].

### 3.2 Semantic Mapping

The basic idea of the semantic mapping, depicted in Fig. 2, is that an "instance" of the activity diagram is represented by some system model concepts such as objects, threads, etc. The abstract names `e1`, `c1` and so on for these entities have been chosen deliberately to not suggest any specific choice. While Fig. 2 highlights only one instance, there may be multiple instances of the same diagram executing concurrently.

For an execution trace and a fixed instance it is then checked if all state transitions $s_{i+1} \in \Delta(s_i)$ conform to the behavior prescribed by the diagram. In Fig. 2, for example, the system model concept that represents action `A` has to be executed prior to the system model concepts that represent actions `B` and `C`. Thus, the inner semantics presented in this Section defines possible orders of executions of actions. How these actions manifest in a system is left open and can be detailed by "fixing" the variation points of the inner semantics.

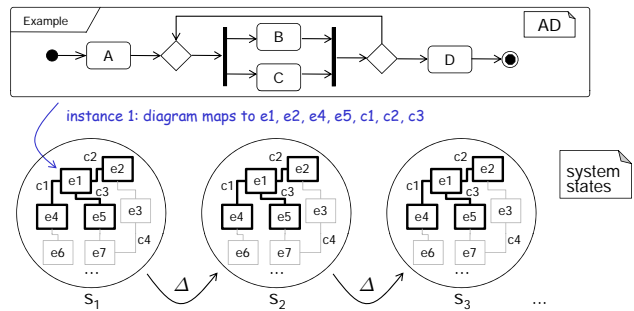

**Fig. 2.** Idea of mapping activity diagrams

The following definitions are given in standard maths. However, according to our approach presented in [8], all definitions (including abstract syntax) will be encoded in a theorem prover to obtain a machine-checkable language definition which is suitable for verification purposes.

In Definition 2, we introduce the set ADInst, i.e., the set of activity diagram instances. Depending on the intended interpretation of the activity diagram, it has to be possible to obtain, e.g., the representation of roles or actions as system model concepts. The exact definition is subject to specific interpretation and is consequently defined as a variation point.

**Definition 2 (Variation point for activity diagram instances).** ADInst *denotes a set of activity diagram instances for an activity diagram. Given an instance, we obtain the corresponding activity diagram by function* ad : ADInst $\to$ AD. *No further assumptions are made on the number and structure of elements of* ADInst *or on function* ad.

Our semantics is completely abstract in terms of how we represent an instance of an activity diagram as entities in the system. Establishing a connection between the *inner* semantics of the diagram and possible realizations is the aim of Sect. 4 where we discuss realization variants.

In each state of the system, information about the currently executing actions is required. Since the mapping of actions to system entities is not fixed, this also remains a variation point.

**Definition 3 (Variation point of state of actions).** *Function* executing : Node $\times$ ADInst $\times$ STATE $\to$ Bool *checks if a given node is currently executing for an instance in a system state.*

Pin types pose a restriction on what tokens may flow into or out of the nodes. This is defined in Definition 4.

**Definition 4 (Variation point for assigning tokens to a pin type).** *Function* elems : PType $\to$ $\wp$(Token) *yields a set of tokens that match the pin type. If the type is the special type* $\top$, *then all tokens are valid (arbitrary data or just control), i.e.,* elems($\top$) = Token. *The special control token* $\bot$ *is the only token matching* $\bot$, *i.e.,* elems($\bot$) = $\{\bot\}$.

To completely capture the current configuration of an activity diagram instance, control and data flow tokens that sit on transitions need also be considered. This is introduced in the following definition. All data are tokens as well. Function bufState in Definition 5 gives access to the current token buffer of a transition in a state. Elements in the buffer have to match the pin types the transition is connected to. These types are not necessarily equal but *compatible*. No further assumptions are made on the behavior of the buffer.

**Definition 5 (Variation point on tokens and token buffers).** Token *is a set of control and data tokens.* bufState : Transition $\times$ ADInst $\times$ STATE $\to$ Buffer(Token) *returns the current buffer of a transition in a state. Given a transition t, instance inst, and state s, the tokens in the buffer match the pin types of the transition:*

$\forall e \in \text{bufState}(t, inst, s) :$
  $e \in (\text{elems}(\text{ad}(inst).\text{pinType}(t.\text{InPin})) \cap \text{elems}(\text{ad}(inst).\text{pinType}(t.\text{OutPin})))$

For convenience, we define bufEmpty$(t, inst, s) = ($bufState$(t, inst, s) = \epsilon)$ and bufNonEmpty$(t, inst, s) = ($bufState$(t, inst, s) \neq \epsilon)$.

Further, we determine the tokens produced and consumed on a transition in a system model step, $s' \in \Delta(s)$ in Definition 6.

**Definition 6 (Consumption and production of tokens).** *The function* cons : Transition $\times$ ADInst $\times$ STATE $\times$ STATE $\to$ Token$^*$ *returns tokens that have been consumed from a transition between to system states. Function* prod : Transition $\times$ ADInst $\times$ STATE $\times$ STATE $\to$ Token$^*$ *yields tokens that have been produced on a transition, respectively.*

Outputting a token may be guarded. We do not specify syntax nor semantics for the language Guard but, according to Definition 7, assume a function that evaluates guards given a context.

**Definition 7 (Variation point on evaluation of guards).** *Function* eval : Guard $\times$ ADInst $\times$ STATE $\to$ Bool *evaluates guards.*

isInitial in Definition 8 checks if a system state corresponds to an initial activity diagram configuration. A system state corresponds to an initial configuration if there are only tokens on the outgoing transitions of initial actions and no other action is currently executing.

**Definition 8 (Initial states of a system).** *For an instance $inst \in$ ADInst, the function* isInitial : ADInst $\times$ STATE $\to$ Bool *determines if a state $s \in$ STATE is an initial state:*

$$\begin{aligned}
&\text{isInitial}(inst, s) = \\
&\quad (\exists n \in \text{ad}(inst).\text{Node} : \\
&\quad\quad (n.\text{NType} = \text{initial} \land \forall t \in \text{outT}(\text{ad}(inst), n) : \text{bufNonEmpty}(t, inst, s))\land \\
&\quad (\forall n \in \text{ad}(inst).\text{Node} : \\
&\quad\quad (n.\text{NType} \neq \text{initial} \implies \forall t \in \text{outT}(\text{ad}(inst), n) : \text{bufEmpty}(t, inst, s)\land \\
&\quad\quad\quad\quad\quad\quad\quad\quad \neg \text{executing}(n, inst, s)))
\end{aligned}$$

A system state corresponds to a final configuration (Definition 9) if there are only tokens on the ingoing transitions of final actions and no other action is executing.

**Definition 9 (Final states of a system).** *For an instance $inst \in$ ADInst, the function* isFinal : ADInst $\times$ STATE $\to$ Bool *determines if a state $s \in$ STATE is a final state:*

$$\begin{aligned}
&\text{isFinal}(inst, s) = \\
&\quad (\exists n \in \text{ad}(inst).\text{Node} : \\
&\quad\quad (n.\text{NType} = \text{final} \land \exists t \in \text{inT}(\text{ad}(inst), n) : \text{bufNonEmpty}(t, inst, s))\land \\
&\quad (\forall n \in \text{ad}(inst).\text{Node} : \\
&\quad\quad (n.\text{NType} \neq \text{final} \implies \forall t \in \text{inT}(\text{ad}(inst), n) : \text{bufEmpty}(t, inst, s))\land \\
&\quad\quad\quad\quad\quad\quad\quad\quad \neg \text{executing}(n, inst, s)))
\end{aligned}$$

Two things may be noted here: a) Requiring that no other action is executing in an initial or final state results in unique, non-overlapping activity diagram instances with respect to system model entities. That means, changing the state in one instance does not affect any other instance. Currently, we still investigate under which conditions overlapping should be admissible since it enables interference between diagram instances which can be desired or unwanted. b) As an extension to Definition 9, we could define some pre-final state condition in that, although a final node was reached, other actions may still execute. Depending on the context, we could allow actions to carry on for some extra time to complete their tasks or kill them immediately.

We now define if a step in the system from state $s$ to state $s'$ with $s' \in \Delta(s)$ by an instance conforms to the behavior prescribed by the activity diagram. This definition may be extended if additional node types (such as hierarchical nodes) are defined.

**Definition 10 (A well behaving system step).** *The function* step *with signature* step : Node $\times$ ADInst $\times$ STATE $\times$ STATE $\to$ Bool *prescribes the allowed behavior in a system step w.r.t. an instance inst according to node $n$.*

$$\text{step}(n, inst, s, s') =$$
$$((n.\text{NType} = \text{action} \implies$$
$$(\text{startAct}(n, inst, s, s') \lor \text{finishAct}(n, inst, s, s') \lor \text{stepInst}(n, inst, s, s'))) \land$$
$$(n.\text{NType} = \text{forkjoin} \implies \text{stepForkJoin}(n, inst, s, s')) \land$$
$$(n.\text{NType} = \text{decisionmerge} \implies \text{stepDecisionMerge}(n, inst, s, s'))$$
$$\lor \text{stutter}(n, inst, s, s'))$$

The following function definitions all have the same signature like step.

**Definition 11 (A stutter step).** *The function* stutter *checks for a stutter step: The execution state (w.r.t. the instance inst) does not change and no tokens are consumed or produced.*

$$\text{stutter}(n, inst, s, s') =$$
$$(\text{executing}(n, inst, s) = \text{executing}(n, inst, s') \land$$
$$\forall t \in \text{inT}(\text{ad}(inst), n) : \text{cons}(t, inst, s, s') = \epsilon \land$$
$$\forall t \in \text{outT}(\text{ad}(inst), n) : \text{prod}(t, inst, s, s') = \epsilon)$$

**Definition 12 (Starting an action node).** *The function* startAct *checks for a start of an action node: execution is started and the required token is consumed.*

$$\text{startAct}(n, inst, s, s') =$$
$$(\neg \text{executing}(n, inst, s) \land \text{executing}(n, inst, s') \land$$
$$(\forall t \in \text{inT}(\text{ad}(inst), n) : \#(\text{cons}(1, t, inst, s, s')) = 1) \land$$
$$(\forall t \in \text{outT}(\text{ad}(inst), n) : \text{prod}(t, inst, s, s') = \epsilon))$$

**Definition 13 (Finishing an action node).** *The function* finishAct *checks for a finishing step of an action node: Execution is stopped and the required token is produced.*

$$\text{finishAct}(n, inst, s, s') =$$
$$(\text{executing}(n, inst, s) \land \neg \text{executing}(n, inst, s') \land$$
$$(\forall t \in \text{outT}(\text{ad}(inst), n) : \#(\text{prod}(t, inst, s, s')) = 1) \land$$
$$(\forall t \in \text{inT}(\text{ad}(inst), n) : \text{cons}(t, inst, s, s') = \epsilon))$$

While startAct and finishAct allow for behavior of nodes that last longer than one system step, stepInst in Definition 14 is appropriate when the execution of a node can be finished in just one step.

**Definition 14 (Instant reaction of an action node).** *The function* stepInst *checks for a step of an action node that constitutes of executing the whole action.*

$$\begin{aligned}
&\text{stepInst}(n, inst, s, s') = \\
&\quad ((\forall t \in \mathsf{inT}(\text{ad}(inst), n) : \#(\text{cons}(t, inst, s, s')) = 1) \land \\
&\quad (\forall t \in \mathsf{outT}(\text{ad}(inst), n) : \#(\text{prod}(t, inst, s, s')) = 1))
\end{aligned}$$

It is assumed that a node produces or consumes at most one token on each transition at a time. Definition 6 allows for a more general treatment where multiple tokens are considered. This can be exploited in future versions of the semantics when considering, for example, streams of tokens and parameter sets [1].

**Definition 15 (Step on a fork/join node).** *The function* stepForkJoin *checks for a step of a fork or join node (or a combination of both): On all input transitions a token is consumed while on all output transitions a token is produced.*

$$\begin{aligned}
&\text{stepForkJoin}(n, inst, s, s') = \\
&\quad ((\forall t \in \mathsf{inT}(\text{ad}(inst), n) : \#(\text{cons}(t, inst, s, s')) = 1) \land \\
&\quad (\forall t \in \mathsf{outT}(\text{ad}(inst), n) : \#(\text{prod}(t, inst, s, s')) = 1))
\end{aligned}$$

Please note, according to Definition 15, the reaction of a fork/join node is instantaneous. While this may be adequate for many interpretations, it may be inappropriate for others. An alternative definition could introduce a two-phase behavior of fork/join similar to that of action nodes. Also a combined definition (instantaneous *or* delayed) is possible. The same holds for Definition 16.

Another interesting issue is the buffering of tokens. There is no need to produce all tokens in one go. What it means to store or retrieve a token depends on how the buffer is "implemented". A rather sophisticated but useful way would be to store arriving data values in attributes (one for each incoming pin) and use an intelligent controller that senses if all tokens arrived (i.e., all attributes are set). This would then be the instant in time at which all tokens are produced.

**Definition 16 (Step on a decision/merge node).** stepDecisionMerge *checks for a step of a decision or merge node (or a combination of both): There is exactly one input token consumed and exactly one output token produced on an output pin with its guard evaluated to true.*

$$\begin{aligned}
&\text{stepDecisionMerge}(n, inst, s, s') = \\
&\quad ((\exists t \in \mathsf{inT}(\text{ad}(inst), n) : \#(\text{cons}(t, inst, s, s')) = 1 \land \\
&\quad\quad \forall t' \in \mathsf{inT}(\text{ad}(inst), n) : t' \neq t \implies \text{cons}(t, inst, s, s') = \epsilon) \land \\
&\quad (\exists t \in \mathsf{outT}(\text{ad}(inst), n) : \\
&\quad\quad \#(\text{prod}(t, inst, s, s')) = 1 \land \text{eval}(\text{ad}(inst).\mathsf{guard}(t.\mathsf{OutPin}), inst, s') \land \\
&\quad\quad \forall t' \in \mathsf{outT}(\text{ad}(inst), n) : t' \neq t \implies \text{prod}(t, inst, s, s') = \epsilon))
\end{aligned}$$

The last definition of the inner semantics of activity diagrams now defines a satisfaction relation of a trace of the system with an activity diagram instance. The conditions that need to be fulfilled are: a) there is a state with an initial configuration ($i$-th state), and all subsequent steps b) behave according to function *step*, or c), if the execution reached a final configuration, it remains final.

**Definition 17 (A trace that satisfies an activity diagram instance).** *For a trace $t \in \text{TRACE}$ of a system in the system model and an activity diagram instance $inst \in \text{ADInst}$, $t$ satisfies $inst$, $t \models inst$, exactly if*

$$(\exists i : \text{isInitial}(inst, t[i]) \land \\ (\forall j \geq i, n \in \text{ad}(inst).\mathsf{Node} : \\ \quad \text{step}(n, inst, t[j], t[j+1]) \land \\ \quad (\text{isFinal}(inst, t[j]) \implies \text{isFinal}(inst, t[j+1])))))$$

Until now, we have not clarified how an activity diagram instance may look like under a specific interpretation. This is the aim of the next section.

## 4 Variants

In this section two variants of activity diagram interpretations are introduced. The degree of formality varies. A rather complete treatment of activity diagrams describing a single method execution which is made up of atomic actions is given. Activity diagrams in which actions are treated as complete methods are discussed informally. Further variants are briefly discussed in the conclusion.

### 4.1 Variant 1: Nodes as atomic actions

Consider the example in Fig. 3. The activity diagram describes an algorithm to compute the factorial of a number. Each action is assumed to be an atomic action. The whole activity is a method definition. The idea now is to specify variants of the definitions in Sect. 3 which are variation points to obtain a semantics in which we interpret an activity diagram as a single method. An instance of the activity diagram hence is a concrete single execution of that method. In this case, an activity diagram instance can be characterized by the following definition:

**Definition 18 (Variant of Definition 2: Activity Diagram Instances).** *An instance of an activity is a single method execution. The following functions constitute the context of the execution:*

- caller : $\text{ADInst} \to \text{UOID}$ *is the caller of the method.*
- meth : $\text{ADInst} \to \text{UMETH}$ *is the method described by the activity diagram.*
- params : $\text{ADInst} \to \text{UVAR}^*$ *is the list of parameters of the method.*
- callee : $\text{ADInst} \to \text{UOID}$ *is the called object. It has to define the method, i.e., $\text{classOf}(\text{callee}(inst)) = \text{definedIn}(\text{meth}(inst))$.*
- pc : $\mathsf{Node} \times \text{ADInst} \to \text{UPC}$ *is a valid program counter value of the action for the specified method, i.e., $\text{pc}(n, inst) \in \text{pcOf}(\text{meth}(inst))$*
- thread : $\text{ADInst} \to \text{UTHREAD}$ *is the thread executing the method.*

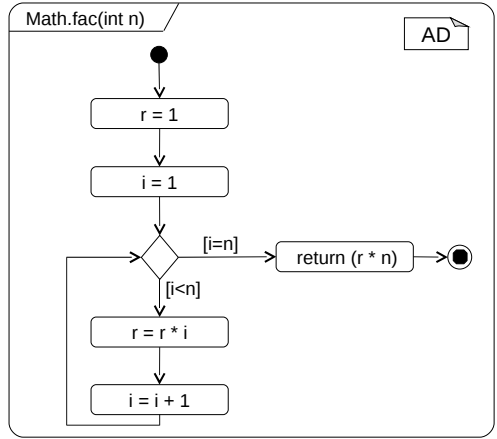

**Fig. 3.** Activity for method "fac"

Since all actions are atomic actions, we need not consider the execution state of a node. All executions are instantaneous and Definition 3 remains underspecified.

No data but only control flow is relevant in this variant, hence we set $\mathsf{PType} = \{\bot\}$ and define $\text{elems}(\bot) = \{\bot = \text{thread}(inst)\}$ for an instance $inst$. This is to model the fact that if there is a token in the buffer of a transition, then the target of the transition is the next action to execute. This is the case if the program counter of the current stack frame identified by some object and thread points to the node which is targeted. For some instance $inst$, transition $t$, and state $s$, this means

$\quad \text{bufState}(t, inst, s) = [\text{thread}(inst)]$
$\quad \Leftrightarrow \text{pcOf}(\text{top}(\text{csOf}(s)(\text{callee}(inst))(\text{thread}(inst)))) = \text{pc}(t.\mathsf{Dst}, inst)$

where top is the first element of the stack. This fixes the variation points of Definitions 5 and 4.

We assume a given action language $\mathsf{AL}$ and set $\mathsf{Effect} = \mathsf{AL}$ and $\mathsf{Guard} \subseteq \mathsf{AL}$. Semantics is traditionally defined: An atomic action is evaluated in the context of an object and a thread that execute an action and it is checked whether the state $s'$ mirrors the effect of executing the action in state $s$. Consider, for example, an action for setting an attribute $\mathsf{SetAttr}\ \mathsf{x}\ \mathsf{y}$: the data store of the object is updated according to the given attribute $x$ and value $y$. Also the program counter is advanced, i.e.,

$\quad \text{sem} : \mathsf{AL} \times \text{UOID} \times \text{UTHREAD} \times \text{STATE} \times \text{STATE} \to \text{Bool}$
$\quad \text{sem}(\mathsf{SetAttr}\ \mathsf{x}\ \mathsf{y}, oid, th, s, s') =$
$\quad\quad (\text{dsOf}(s')(oid) = \text{dsOf}(s)(oid) \oplus [x \mapsto y] \wedge$
$\quad\quad \text{csOf}(s')(oid)(th) = \text{incPC}(\text{csOf}(s)(oid)(th)))$

In order to make sure that actions are properly executed, we add the constraint that executing a node $n$ in instance $inst$ corresponds to considering its effect:

$\quad \text{stepInst}(n, inst, s, s') \Leftrightarrow \text{sem}(n.\mathsf{Effect}, \text{callee}(inst), \text{thread}(inst), s, s')$

Decision nodes determine the next action to execute based on their guards. Since this is done by setting the program counter to the right value, guards have side effects in this variant. In order to reflect this, we complement Definition 16 with:

$$\text{stepDecisionMerge}(n, inst, s, s')$$
$$\Leftrightarrow \exists t \in \text{outT}(\text{ad}(inst), n) :$$
$$\text{sem}(\text{ad}(inst).\text{guard}(t.\text{OutPin}), \text{callee}(inst), \text{thread}(inst), s, s')$$

To not contradict Definition 16, we assume $\text{eval}(g, inst, s)$ to hold for all guards, instances, and transitions. Consequently, Definition 16 just ensures that exactly one token is consumed and produced while the above constraint ensures that the effect of a guard was observed in a system step.

*Syntactic consequences:* According to this variant and the semantics defined for it, fork/join nodes should be excluded syntactically since there is only one thread or token. The sequential execution by one thread also indicates that all nodes (except for decision nodes) may have only one output pin and that roles are excluded as well. Since only control flow between atomic actions is modeled, data types on pins are also disallowed.

### 4.2 Variant 2: Actions as methods

In this variant, all actions are considered to be complete methods of some objects instead of atomic actions of one method. This might be a suitable interpretation of the activity diagram in Fig. 1.

Activity diagram instances can in this case be characterized as in Definition 19: Nodes correspond to methods, there is a set of threads executing these methods. A specific object on which the method is called is obtained by oid. Roles are represented as objects as well.

**Definition 19 (Variant of Definition 2: Activity Diagram Instances).** *An instance of an activity diagram in which actions denote methods is characterized by the following functions:*

- meth : Node × ADInst → UMETH *is the method referenced by the node.*
- threads : ADInst → $\wp$(UTHREAD) *is the set of threads in that instance.*
- oid : Node × ADInst → UOID *is an object that holds a method for the node.*
- rrep : Role × ADInst → UOID *is the object representing the role.*

Instances can be refined further by introducing sub-variants of Definition 19. For example, we may require that for an instance $inst$, the role of node $n$ is defining the method, i.e.,

$$\text{definedIn}(\text{meth}(n, inst)) = \text{classOf}(\text{rrep}(\text{ad}(inst).\text{roleOf}(n), inst))$$
$$= \text{classOf}(\text{oid}(n, inst))$$

According to Fig. 1, for example,

$$\text{definedIn}(\text{meth}(\textsf{Evaluate}, inst)) = \text{classOf}(\text{rrep}(\textsf{Referee1}, inst))$$

So the method that implements action Evaluate is defined in a class that represents role Referee1. An interesting question in this context then is: Who is calling an action (i.e., method). Is it done by the role itself? Is there some additional control structure that checks if a role finished one of its methods and then calls (enables) the next one? Methods may, however, not be associated to roles at all. Instead, specific objects, structured roughly as in the *command* design pattern [9] could represent action nodes. To ensure data integrity, one has to be careful when allowing concurrent instances, i.e., concurrent executions of methods (or of a single method as in variant 1). Analysis of the activity diagrams would be required to proof or refute this property. At this state, we cannot faithfully give definite answers but will examine these questions in future work.

Executing a node (cf. Definition 3) means executing a method, so there has to be a stack frame $f = (\mathrm{oid}(n, inst), \mathrm{nameOf}(\mathrm{meth}(inst)), *, *, *)^4$ for a thread $th \in \mathrm{threads}(inst)$, i.e.,

$$\mathrm{executing}(n, inst, s) \Leftrightarrow f \in (\mathrm{csOf}(s)(\mathrm{oid}(n, inst))(th))$$

Again, we could be more specific. For example, we could force the caller of the method to be the object that is representing the role, i.e., we have the last component of $f$ equal to $\mathrm{rrep}(\mathrm{ad}(inst).\mathsf{roleOf}(n), inst)$. In this variant, a natural interpretation of tokens are method calls in the system. However, there is again more than one choice. A token may correspond to a call that is both carrying control and data. A single token could then correspond to a possibly complex parameter list for the method. In case of multiple incoming transitions to a node, we could also follow the idea discussed earlier that the actual method execution can only be started if all required input data has arrived on all input pins. The (incomplete) code snippet in Fig. 4 informally shows a possible implementation in which an action `HandlePerson` is mapped to a class which has attributes for all input pins. Setting the attribute also informs some controller that keeps track of the state of attributes. Once all attributes are set, the controller may call method `exec` that implements the actual behavior. Syntactically, all features introduced in Sect. 2 make sense in this variant, so there are no syntax restrictions as in the previous variant.

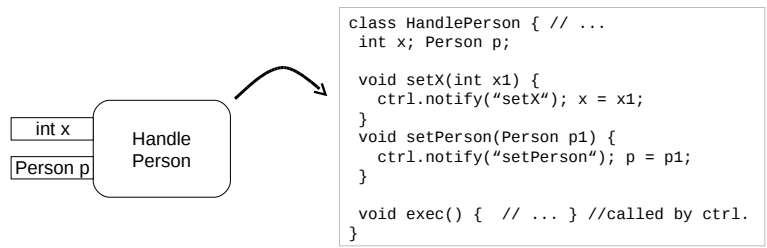

**Fig. 4.** Special buffering strategy for tokens as attributes

---

[4] Values we are not interested in can be marked as "wild card" by $*$

## 5 Related Work

The common denominator of most works regarding the semantics of activity diagrams is the idea to define the possible orders of executions of actions. In that respect, our semantics is not different. The UML standard defines an informal token flow semantics with semantic variation points [1]. However, the standard provides no means to describe realizations. In our approach, we obtain realizations by stating variants of several function definitions. A formal approach uses procedural Petri-nets for the semantics of UML2 activity diagrams [10]. Here, only the control flow aspect of activity diagrams is covered (including concurrency and procedure calls), whereas data flow is covered in our approach as well. As an extension, the data flow in activity diagrams has been mapped to Colored Petri-nets [11]. Both works do not consider a specific application domain. Eshuis [4] develops a requirements-level and an implementation-level semantics for activity diagrams. Both semantics are fixed and focus on workflow management systems while we introduce an inner semantics from which variants can be developed. Another token-based approach in the application area of workflow management systems uses a virtual machine to execute activities [12]. Here, a fixed semantics is defined by mapping a model to its execution in said runtime engine. The semantics of UML actions is formally defined using the system model as a semantic domain in [13]. An extension of this work [14] describes a virtual machine for UML2 actions and activities based on a fixed interpretation in the system model. Another natural candidate to formalize the semantics of activities are process calculi. For example, in [15] the $\mu$-calculus is used. The proposed Petri-net and process calculus semantics often have the advantage of being executable and analyzable but do not allow an easy understanding of models in terms of possible implementations.

## 6 Conclusion

We have defined a formal semantics for a subset of UML activity diagrams. The inner semantics was equipped with variation points that can be interpreted differently in specific application domains. Variants are obtained by deciding which system model entities make up a diagram instance and how their execution state and token flow is determined. This was sketched using two example variants.

Having clarified the inner semantics of activity diagrams in terms of the system model, we are now working towards formalizing different variants of activity diagram interpretations. In this paper, we were mainly concerned with rather low-level interpretations of activity diagrams as simple action or method executions. As discussed in [3], activity diagrams can be used to model simple web page flows but also complex collaborations in web information systems. Executing an action in this context means, for example, showing a web page to a user, waiting for a data update, and storing it in a session context or data base. Another interesting line of future work is to include further concepts from activity diagrams, for example, interruptible activity regions, parameter sets, etc. However, there is the danger of cluttering the notation with constructs which are only useful in very special situations. To avoid this, we will introduce these concepts as syntactic variants in addition to a relatively small language core as explained in [16]. Further, we are confident that it is possible to combine different interpretations of activity diagrams when considering hierarchical decomposition. For example, it is possible to model the content of nodes interpreted as methods by diagrams in which nodes are interpreted as basic actions and to adopt the diagram instance.